      [1999/12/01 v1.4c Il Nuovo Cimento]
\documentclass{cimento}

\newcommand{\be}{\begin{equation}} 
\newcommand{\ee}{\end{equation}} 
\newcommand{\ba}{\begin{eqnarray}} 
\newcommand{\ea}{\end{eqnarray}}

\usepackage{graphicx}  
\title{Electroweak corrections to squark--anti-squark pair production at the LHC}
\author{W.~Hollik\from{mpi}\ETC,
E.~Mirabella\from{mpi}}
\instlist{\inst{mpi} Max-Planck-Institut f\"ur Physik (Werner Heisenberg
  Institut) F\"ohringer 
Ring 6, D-80805 M\"unchen, Germany}
\begin{document}

\maketitle

\vspace*{-6cm}
\begin{flushright} 
\small{MPP-2008-53}
\end{flushright}
\vspace{6cm}

\begin{abstract}
We present the complete NLO electroweak contribution to the production of diagonal
squark--anti-squark pairs in proton--proton collisions. 
We discuss their effects for the production of squarks
different from top squarks, in the SPS1a$'$ scenario.
\end{abstract}

\section{Introduction}
TeV-scale supersymmetry (SUSY) will be accessible to
direct experimental studies at the LHC through the production of SUSY
particles.
In particular, colored particles 
will be copiously produced, and the hadronic production of
squark--anti-squark pairs is expected to play an important role 
for SUSY hunting. \\
\noindent
The first prediction of the 
cross section for the process $P~P \to \tilde{Q}^a~\tilde{Q}^{a*}$   
was done at lowest order $\mathcal{O}(\alpha_s^2)$
in supersymmetric QCD~\cite{Tree}. The dominant NLO corrections, of $\mathcal{O}(\alpha^3_s)$,
were calculated more than ten years later~\cite{Beenakker1996}. \\
\noindent
There are also partonic processes of electroweak 
origin, like  diagonal and non-diagonal squark pair production form $q \bar{q}$ annihilation~\cite{Drees}, contributing at  $\mathcal{O}(\alpha_s\alpha)$
and $\mathcal{O}(\alpha^2)$ at the tree-level. In particular  
the interference between the tree-level QCD 
and electroweak amplitudes at $\mathcal{O}(\alpha_s\alpha)$ for $\tilde{Q}\neq \tilde{t}$
can become sizable. \\
\noindent
NLO electroweak (EW) contributions were found to be significant
in the case of top-squark pair production, with effects up to 20\%~\cite{StopEW}. In the  
case of the  production of diagonal squark--anti-squark pair different from the top-squark
\be
\label{Eq:Process}
 P~P \to \tilde{Q}^a~\tilde{Q}^{a*}\, X\quad (\tilde{Q}\neq \tilde{t}) \, ,\,
\ee 
NLO EW corrections can reach the same size as the tree-level EW contributions of $\mathcal{O}(\alpha_s\alpha)$
and $\mathcal{O}(\alpha^2)$~\cite{SquarkEW}.

\section{EW contributions}
Diagrams and  corresponding amplitudes for the EW contributions to the process~(\ref{Eq:Process})
are generated using \verb|FeynArts|, \verb|FormCalc| and \verb|LoopTools|~\cite{Thomas}.
IR and Collinear singularities are regularized within mass regularization.  

\subsection{Tree level EW contributions}
Tree-level EW contributions to the process ~(\ref{Eq:Process}),  are of $\mathcal{O}(\alpha_s \alpha)$ 
and $\mathcal{O}(\alpha^2)$. The interference of the tree-level electroweak and tree-level QCD diagrams  give rise to terms of 
order $\mathcal{O}(\alpha_s\alpha)$, while $\mathcal{O}(\alpha^2)$ terms are obtained squaring the aforementioned tree-level EW graphs. On top, 
we have also the photon-induced partonic process  $\gamma g \to \tilde{Q}^a\tilde{Q}^{a *}$, which contributes at $\mathcal{O}(\alpha_s \alpha)$ owing to the 
non-zero photon density in the proton.

\subsection{NLO EW contributions}
NLO EW corrections arise from two different channels, gluon fusion and quark--anti-quark  annihilation channels. \vspace{0.1cm} \\
\noindent
\underline{Virtual Corrections}.
Virtual corrections arise from the interference of the tree-level diagrams with the one-loop EW graphs.
In the case of $q \bar{q}$ annihilation channels, there is also  the interference between tree-level EW diagrams and QCD graphs.
The renormalization of the squark, of the quarks, and of the gluino masses and wavefunctions  
has been performed in the on-shell scheme~\cite{HeidiHollik, DenneHab} while 
the strong coupling is renormalized in the 
$\overline{\mbox{MS}}$ scheme. Massive particles (top, squarks, and gluino) have been  decoupled subtracting their contribution at zero momentum transfer. 
Dimensional regularization spoils SUSY at higher order; we restore it by adding a finite counterterm for the renormalization of the
$\tilde{q}\tilde{g}\bar{q}$ Yukawa coupling. \vspace{0.1cm} \\
\noindent
\underline{Real Corrections}.
In order to obtain  IR and collinear finite results we need to include 
the processes of real photon emission. In the case of  $q \bar{q}$ annihilation 
real gluon emission  of $\mathcal{O}(\alpha^2_s \alpha)$ has to be considered as well. 
The photon momentum integration and isolation
of divergences has been performed using two different methods: phase space slicing and dipole subtraction. They give results in good numerical 
agreement. \\
\noindent 
IR singularities drop out  in the sum of virtual and real corrections. Collinear singularities do not and have to be absorbed into the definition 
of parton distribution functions (PDF) of the quarks.

\section{Numerical Results}
\begin{figure}
\centering
\underline{$\tilde{u}^{R}$}\\
\includegraphics[width=6.5cm]{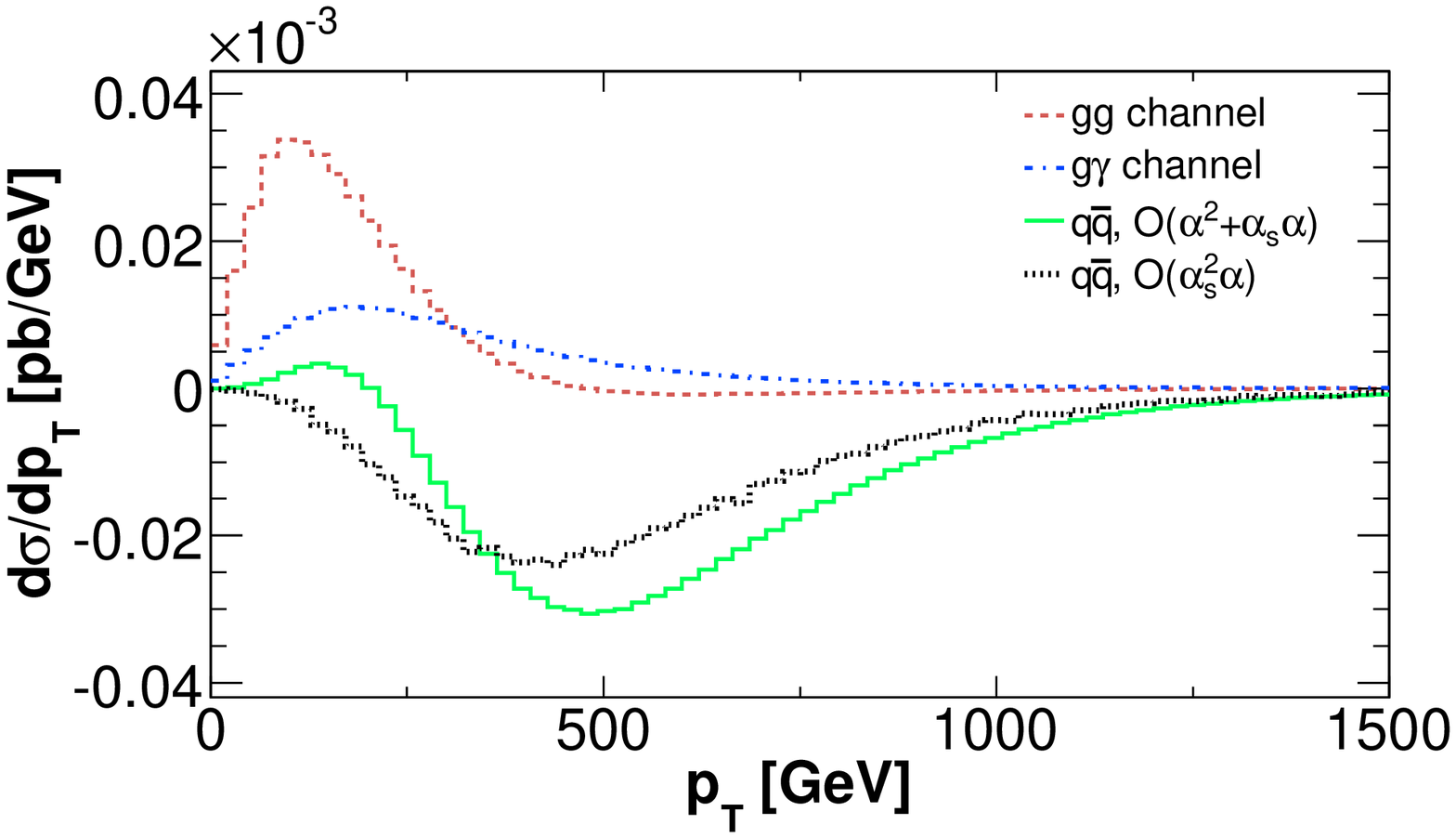}
\includegraphics[width=6.5cm]{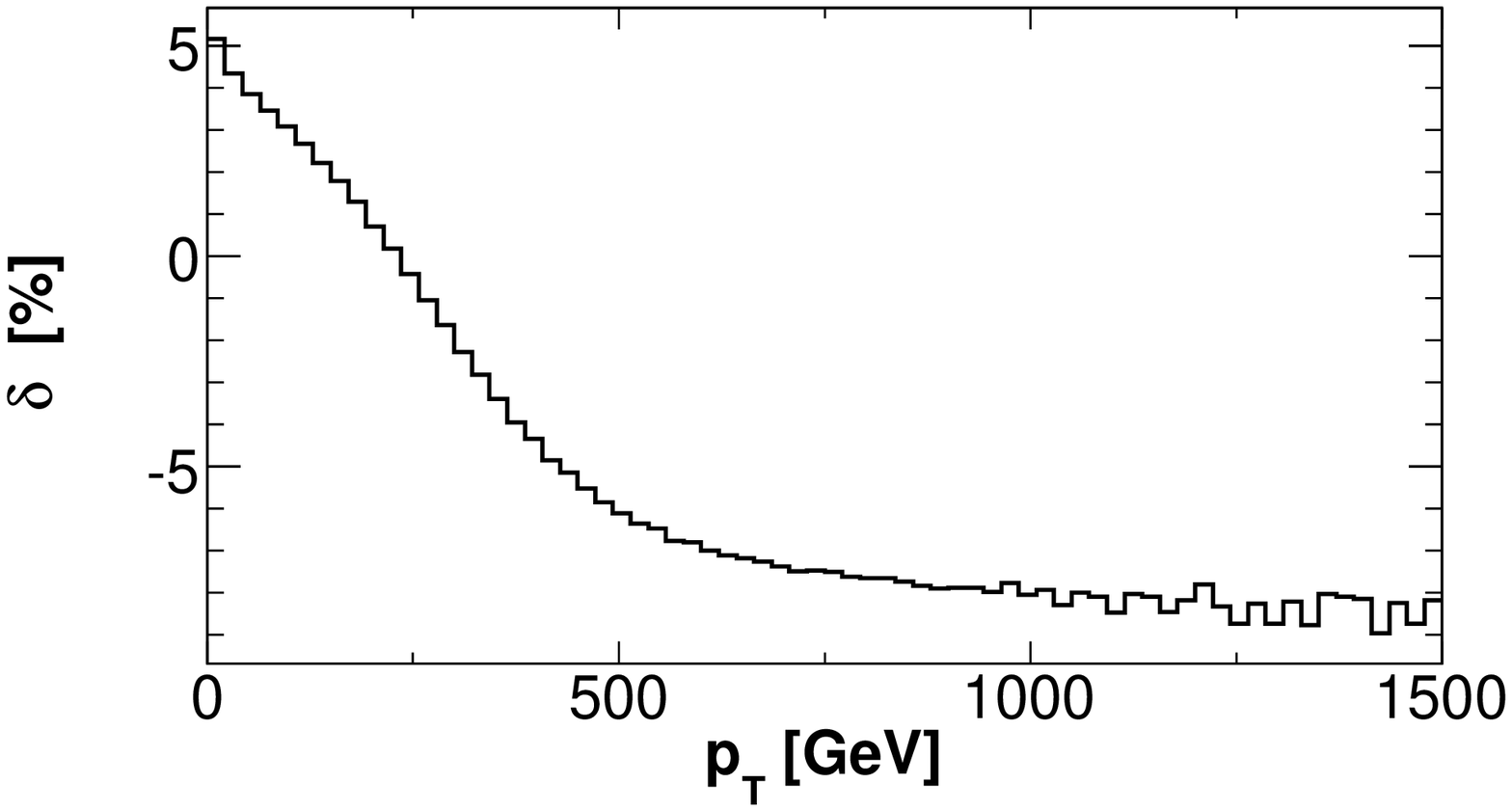}
\centering
\underline{$\tilde{u}^{L}$}\\
\includegraphics[width=6.5cm]{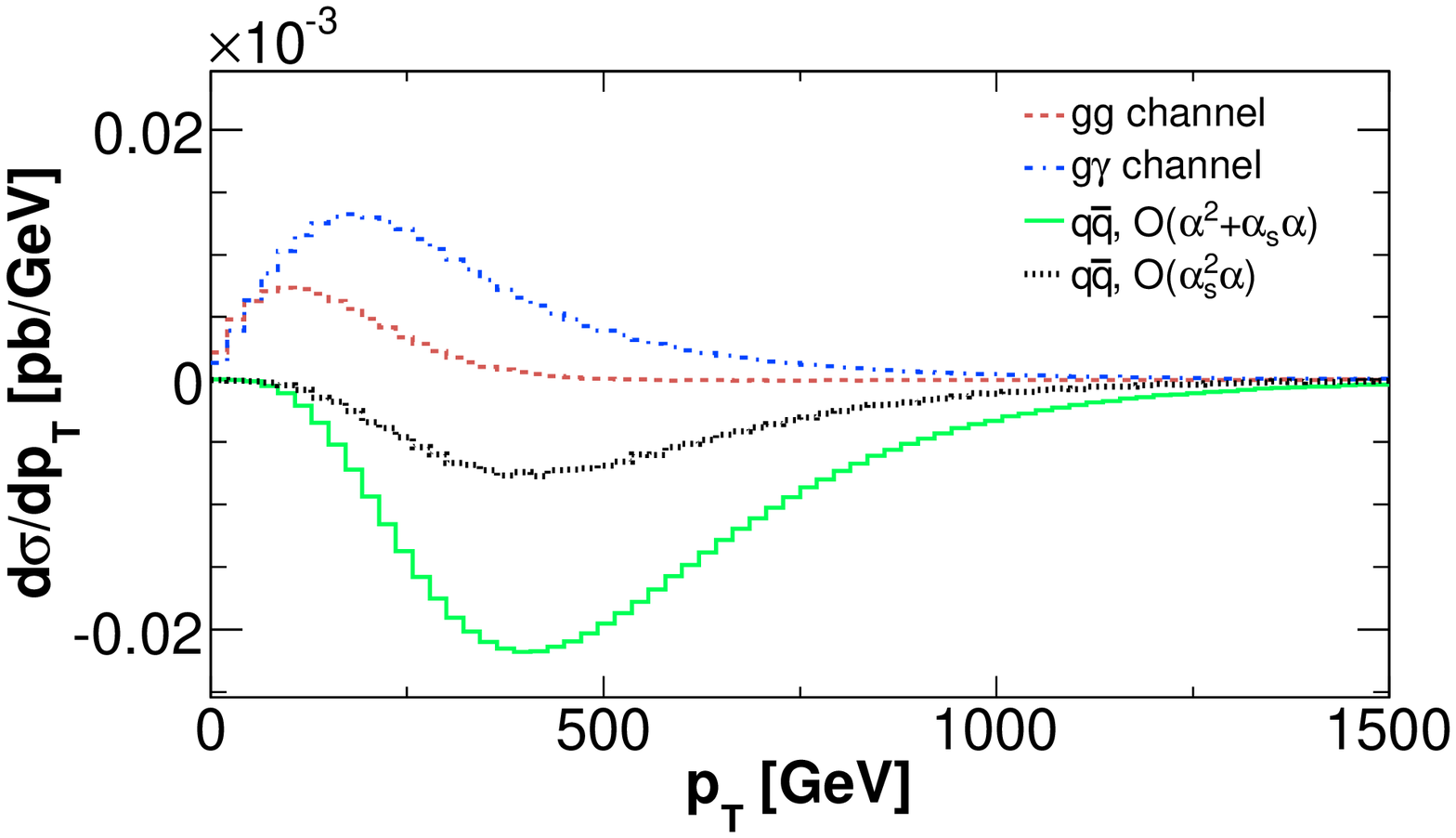}
\includegraphics[width=6.5cm]{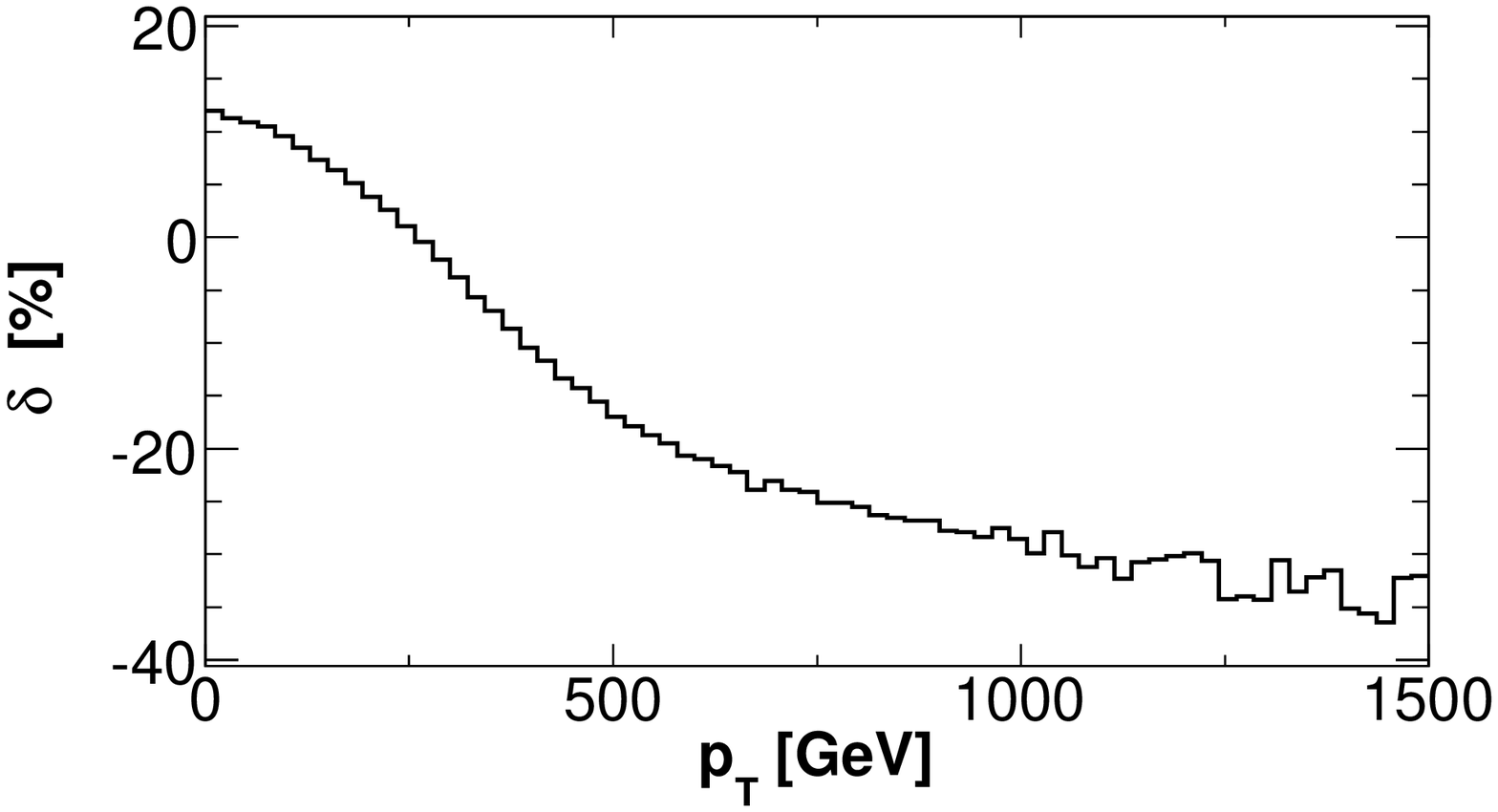}
\caption{Transverse momentum distribution for $\tilde{u}^L \tilde{u}^{L*}$ and
$\tilde{u}^R \tilde{u}^{R*}$ production
for the SUSY parameter point corresponding to SPS1a$'$.
The left panels show the contributions from the various channels.
$\delta$ is the total  EW contribution relative
to the LO one.}
\label{Fig01}
\end{figure}

For illustration of the EW effects, we study the production of the
two up-squark  $\tilde{u}^R$ and $\tilde{u}^L$ focusing on the SPS1a$'$ point of the MSSM parameter space, as 
suggested by the SPA convention~\cite{SPA}. A more comprehensive analysis
can be found in ref.~\cite{SquarkEW}. \\
\noindent
Fig.~\ref{Fig01} contains the transverse momentum distribution of the squarks.
The contribution from the gluon fusion channel is always positive and dominates at lower values of $p_T$, wheras the 
$q \overline{q}$ annihilation channel part is negative and renders the EW contribution negative in the limit of large $p_T$.
The contribution of the $g \gamma$ channel is independent on the squark chirality, determined
only by the electric charge of the produced squarks. \\
\noindent
The EW effects are more pronounced for left-handed chirality
yielding more than 30\% negative contributions for large $p_T$. In the $\tilde{u}^L$ case the LO EW contribution are positive for low $p_T$, 
originating from the PDF-enhanced parton process $d\bar{d} \rightarrow \tilde{u}^L\tilde{u}^{L *}$
through $t$-channel chargino exchange. This positive part is practically compensated by the NLO $\mathcal{O}(\alpha_s^2\alpha)$
contributions in the $q\bar{q}$ annihilation channels. \\

\end{document}